\documentclass[aps,
pra,
twocolumn,
nofootinbib,
amsmath,
amssymb,
superscriptaddress,
reprint,
10pt
]{revtex4-2}

\usepackage[hyperindex,colorlinks,breaklinks,allcolors=blue]{hyperref}

\usepackage{amsmath, amssymb}
\usepackage{graphicx}
\usepackage{tikz}
\usepackage[utf8x]{inputenc}
\usepackage{color}
\usepackage{bm}

\usepackage{pstool}
\usepackage{pgf}
\usepackage{import}

\usepackage{cleveref}

\usepackage{multirow}
\usepackage{array}

\usepackage{relsize}
\usepackage{xspace}

\usepackage{todonotes}

\newcommand{\del}{\partial}

\newcommand{\vev}[1]{\langle #1 \rangle}

\newcommand{\eg}{\textit{e.\,g.}}
\newcommand{\td}{\text{d}}
\newcommand{\zh}{z_\text{h}}

\newcommand{\Tc}{T_c}
\newcommand{\Phit}{\tilde\Phi}

\newcommand{\Rplus}{\protect\hspace{-.1em}\protect\raisebox{.35ex}{\smaller{\smaller\textbf{+}}}}
\newcommand{\Cpp}{\mbox{C\Rplus\Rplus}\xspace}

\definecolor{ao(english)}{rgb}{0.0, 0.5, 0.0}
\definecolor{applegreen}{rgb}{0.55, 0.71, 0.0}
\definecolor{cadetblue}{rgb}{0.37, 0.62, 0.63}
\definecolor{cadet}{rgb}{0.33, 0.41, 0.47}
\definecolor{byzantine}{rgb}{0.74, 0.2, 0.64}
\definecolor{orange}{rgb}{1.0, 0.5, 0.0}

\newcommand{\ie}{\textit{i.\,e.}}
\newcommand{\cf}{\textit{cf.}}

\renewcommand{\i}{\mathrm{i}}

\makeatletter
\let\cat@comma@active\@empty
\makeatother

\begin{document}
	

\title{Quantum Turbulence in a Three-Dimensional Holographic Superfluid}

\author{Paul Wittmer}
\affiliation{Institut f\"ur Theoretische Physik, 
	Ruprecht-Karls-Universit\"at Heidelberg,
	Philosophenweg 16, 
	69120 Heidelberg, Germany}
\affiliation{ExtreMe Matter Institute EMMI, 
	GSI Helmholtzzentrum f\"ur Schwerionenforschung,\\
	Planckstra{\ss}e 1, 
	64291 Darmstadt, Germany}	

\author{Carlo Ewerz}	
\affiliation{Institut f\"ur Theoretische Physik, 
	Ruprecht-Karls-Universit\"at Heidelberg,
	Philosophenweg 16, 
	69120 Heidelberg, Germany}
\affiliation{ExtreMe Matter Institute EMMI, 
	GSI Helmholtzzentrum f\"ur Schwerionenforschung,\\
	Planckstra{\ss}e 1, 
	64291 Darmstadt, Germany}        
	
\begin{abstract}
We report real-time simulations of far-from-equilibrium dynamics of a holographic superfluid 
in three dimensions. The holographic duality maps a strongly coupled superfluid to a
weakly coupled theory with gravity in a higher-dimensional curved space which we study numerically.
Starting from configurations of dense tangles of quantized vortex lines, the superfluid exhibits turbulent behavior during
its evolution towards thermal equilibrium. Specifically, we observe two subsequent
universal regimes of quasi-stationary scaling in the occupation number spectrum, the first of which is  
Kolmogorov-like.
Indications for the strongly dissipative nature of vortex dynamics in the three-dimensional holographic superfluid
are found in the rapid shrinking and annihilation of small
vortex rings that emerge from frequent reconnections of vortex lines during the evolution. 
\end{abstract}
	
	
\maketitle

%
\textit{Introduction.} 
Understanding turbulence in superfluids poses a challenge that has 
attracted great interest over the past decades
\cite{Barenghi2001lecturenotes,Halperin2008a,Tsatsos:2016physrep},
prompted in particular by the experimental observation of Kolmogorov scaling in 
the energy spectrum of superfluid helium \cite{Maurer1998,Stalp1999,Walmsley2014pnas}
and of quantum turbulence in atomic Bose--Einstein condensate (BECs) \cite{Henn:2009turbprl}. 
Superfluid or quantum turbulence is characterized by the presence of quantized vortex defects,
and in three dimensions associated with the presence of a tangle of vortex lines. 
Reconnections of quantized vortex lines, first anticipated in \cite{Feynman1955}, are 
crucial for the energy transfer between different length scales. 
Experimentally, superfluid vortex dynamics has been intensely studied
\cite{Vinen1961,Walmsley2007,Bradley2007,Bradley2011,Bewley2008,Serafini2015,Paoletti2008,Bulgac2013,Serafini2016,Tang2022,Svancara:2023yrf}, 
including the first observation of vortex reconnections in both superfluid
helium \cite{Bewley2008} and BECs \cite{Serafini2015}. 

The theoretical description of the non-equilibrium evolution of superfluids is very intricate
due to strong correlations and nonlinear excitations. Various approaches have been developed
and applied to quantum vortex dynamics and turbulence, see
\cite{Hall1956a,Iordansky1964,Hall1956b,Bekarevich1961,Gross:GPE,Pitaevskii:GPE,Proukakis2008,Schwarz1985,Schwarz1988,Koplik1993,deWaele1994,Gabbay1998,Leadbeater2001,Ogawa2002,Nazarenko2003,Tebbs2010,Nemirovskii2014,Galantucci2018,Vinen2002,Berloff:PRA2002,Berloff2007,Nemirovskii1998,Nemirovskii2002,Berdichevsky2002a,Berdichevsky2002b,Nore1997a,Nore1997b,Araki2002,Kobayashi2002,Tsubota2010,Nemirovskii2013,Kivotides2014,Nowak:2010tm,Nowak:2011sk,Nowak:2012gd,Mathey:2014xxa}.  
Prominent among them are the Gross--Pitaevskii (GP) equation \cite{Gross:GPE,Pitaevskii:GPE}
for the macroscopic wave function of dilute Bose gases and
its extensions for finite-temperature dissipation \cite{Proukakis2008}. 
However, these are not applicable to dense and strongly interacting Bose condensates  
like superfluid helium which exhibit strong dissipation induced by friction
between the defects and thermal excitations.
Other approaches can be used for dense and strongly interacting superfluids but
cannot intrinsically describe the microscopic dynamics of vortex density profiles and of vortex
reconnections, for which they require \textit{ad hoc} assumptions. 

Gauge/gravity duality or holography \cite{Maldacena:1997re,Gubser:1998bc,Witten:1998qj} 
offers an inherently nonperturbative 
description of strongly correlated quantum theories in terms of 
classical gravitational models in a higher-dimensional black-hole spacetime. 
In particular, the classical model intrinsically captures even strongly coupled and 
nonlinear dynamics of the dual quantum theory. 
Applications of holography range across various fields 
\cite{Hartnoll:2009sz,McGreevy:2009xe,Zaanen:2015oix,Ammon:2015wua,Natsuume:2014sfa,Hartnoll:2016apf}. 
Among them, the holographic description of superfluids 
\cite{Gubser:2008px,Hartnoll:2008vx,Herzog:2008he} 
has attracted particular attention, and important findings include 
\cite{Hartnoll:2008kx,Horowitz:2008bn,Albash:2009iq,Montull:2009fe,Keranen:2009re,Arean:2010zw,Arean:2021tks}. 
The holographic description covers all relevant length scales of the superfluid 
in terms of a microscopic dual theory with gravity. This includes the dynamics of topological
defects and their interactions, from vortex cores and their dynamical reconnections
to the infrared. 

For the $(2+1)$-dimensional superfluid, real-time simulations of the dual holographic theory
have been performed to study various aspects of vortex and non-equilibrium dynamics, 
see \eg\ \cite{Adams:2012pj,Ewerz:2014tua,Du:2014lwa,Wittmer:2020mnm,Ewerz:2020wyp,Yang:2022foe}.
Intriguingly, vortex dynamics in the two-dimensional holographic superfluid exhibits dissipation in the
range of real-world superfluids \cite{Wittmer:2020mnm}. It is natural to expect a similarly strong dissipation for
vortex defects in the three-dimensional holographic superfluid. 

Here we employ holography to investigate the non-equilibrium time evolution
of a strongly coupled $(3+1)$-dimensional superfluid by numerical real-time simulations. 
In the following we present our main findings, in large part based on \cite{Wittmer:2021oxf}.  
Technical details and supplemental information are presented in appendices.

%
\textit{Superfluidity} is associated with the 
spontaneous breaking of a global $U(1)$ symmetry in a quantum field theory.
The system is described in terms of the vacuum expectation 
value of a scalar operator,   
$\langle\Psi(t,\bm{x})\rangle = \psi(t,\bm{x})= \sqrt{n(t,\bm{x})}\exp{\{\text{i} \,\varphi(t,\bm{x})\}}$, the superfluid order parameter.
$n(t,\bm{x})=|\psi(t,\bm{x})|^2$ is the superfluid density, and we denote its value in thermal equilibrium by $n_0$. 
The phase $\varphi(t,\bm{x})$ determines the velocity or flow field of the superfluid, $\bm{v}(t, \bm{x})=\bm{\nabla}\varphi(t,\bm{x})$. 
We have three spatial coordinates, $\bm{x}=(x_1,x_2,x_3)$, with corresponding gradient $\bm{\nabla}$. 

%
\textit{In holography}, a superfluid 
in $(3+1)$ dimensions has in the simplest bottom-up construction a dual  
description as an Abelian Higgs model in a $(4+1)$-dimensional asymptotically
anti-de Sitter (AdS) spacetime \cite{Gubser:2008px, Herzog:2008he,Hartnoll:2008vx},
\begin{align} \label{eq:ActionAdSCFT}
S &= \frac{1}{16\pi G_\text{N}^{(5)}} \int d^5 x \sqrt{-\text{det}\,g_{\mu\nu}}\left (\mathcal{R} - 2 \Lambda + \frac 1 {q^2} \mathcal{L}_{\mathrm{gm}} \right), \nonumber \\
\mathcal{L}_{\mathrm{gm}} &= - \frac{1}{4} F_{\mu \nu} F^{\mu \nu}- \lvert D_{\mu} \Phi \rvert^2 - m^2 \lvert \Phi \rvert^2\,.
\end{align}
$G_\text{N}^{(5)}$ is Newton's constant in five dimensions, $g_{\mu \nu}$ the 
spacetime metric with Ricci scalar $\mathcal{R}$, 
and $\Lambda=-6/L_\text{AdS}^2$ the cosmological 
constant of the AdS spacetime with curvature radius $L_\text{AdS}$.
$\mu,\nu=t,\bm{x},z$ with the superfluid's coordinates $t,\bm{x}$
and the holographic coordinate $z$. 
The scalar field $\Phi$ of mass $m$ and charge $q$ is  
coupled to the gauge field $A_\mu$ via the gauge covariant 
derivative $D_\mu=\nabla_\mu - \text{i}\,A_\mu$ with field strength tensor 
$F_{\mu\nu}=-\text{i}\,[D_\mu, D_\nu]$. 

We consider the model \eqref{eq:ActionAdSCFT} in the probe approximation, thus neglecting 
the backreaction of the gauge--matter fields onto the gravitational sector. 
This is valid if the temperature of the system 
is of the order of but below the phase-transition temperature. 
Solving only the gravity part of the 
action \eqref{eq:ActionAdSCFT} yields the `bulk' metric 
\begin{align}\label{eq:Metric}
\text{d} s^2=\frac{L_\text{AdS}^2}{z^2}\left[-\left(1-\frac{z^4}{z_\text{h}^4}\right)\text{d} t^2  +  \text{d}\bm{x}^2-2\text{d} t\,\text{d} z\right]
\end{align}
of the Schwarzschild--AdS spacetime with a planar black hole horizon located at $z=\zh$.
The gauge--matter sector is solved in this fixed background.
The superfluid dynamics is encoded in the near-boundary 
behavior of the scalar-field solution $\Phi$ which reads 
\begin{align} \label{eq:BoundaryExpansion}
\Phi(t,\bm{x}, z)=\eta(t,\bm{x})\,z +  \psi(t,\bm{x}) \,z^3 +\mathcal{O}(z^4) 
\end{align}
for our choice $m^2=-3 L_\text{AdS}^2$. 
Here $\psi(t,\bm{x})$ is the superfluid order parameter 
and $\eta(t, \bm{x})$ is set to zero by appropriate boundary conditions.  
The superfluid has temperature $T=(\pi z_\text{h})^{-1}$ and 
a non-zero chemical potential $\mu$ for the $U(1)$ charge. $\mu$ is fixed by the boundary 
condition of the temporal gauge-field component, $A_t(t,\bm{x},z=0)=\mu$.
The thermodynamic state of the system is controlled by $\mu/T$. 
If $\mu/T$ exceeds a critical value, a scalar charge cloud builds up in the bulk,
corresponding to a second-order phase transition 
into the superfluid state with $\psi\neq 0$ \cite{Horowitz:2008bn}.
We fix our units by setting $z_\text{h}=1$, implying a critical chemical potential of $\mu_c \simeq 4.1568$. 
A choice of $\mu$ then fixes $T/\Tc$.
In this work we choose $\mu=5$, corresponding to $T/\Tc=0.83$. 
The equations of motion and the boundary conditions of the holographic superfluid are discussed in detail in appendix \ref{app:HoloSetup}.

In the probe approximation the black hole effectively acts as an infinitely sized
static heat bath into which energy can dissipate. 
In light of the Tisza--Landau two-fluid model \cite{Tisza1938,Landau1941} of superfluidity,
the black hole may loosely be interpreted as the normal component
while the gauge--matter sector represents the superfluid component.
For the two-dimensional case, this analogy was
established for the backreacted system \cite{Sonner:2010yx}. 

\textit{Quantized vortex lines} are topological defects in the superfluid density
\cite{Onsager1949}, \cite{Feynman1955}. 
Along closed paths encircling the vortex core once, the phase of the order-parameter
field winds by integer $w \neq 0$ multiples of $2\pi$.
At the vortex center the superfluid density vanishes. 
Vortex lines need not be straight and their ends 
can coalesce to form closed loops (vortex rings) \cite{Donnelly1991a}.

\begin{figure}
	\includegraphics[width=\columnwidth]{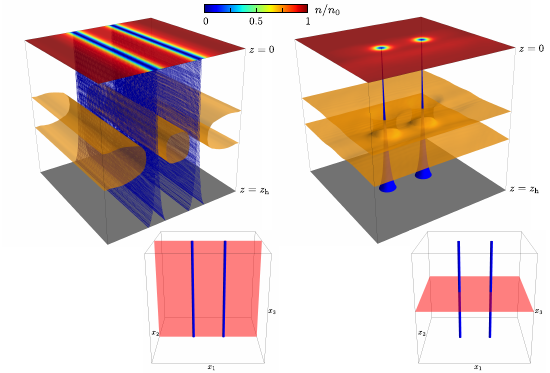}
	\caption{%
		Geometric bulk representation of vortex line solutions outgoing from two two-dimensional 
		slices of the three-dimensional superfluid intersecting the lines as indicated in the small diagrams.  
		Shown are bulk isosurfaces of $|\Phi|^2/z^6$ (blue)
		and of the scalar charge density (orange). 
		The field $|\Phi|^2/z^6$ reduces to the superfluid density $n =\lvert \psi \vert^2 $ 
		at $z=0$ where the color coding refers to $n/n_0$. 
		The gray area at $z=\zh$ represents the black-hole horizon.
	\label{fig:LinesCheese}
              }
\end{figure}

\textit{Holographic bulk views}
of numerical solutions for two anti-parallel straight vortex lines are shown in Fig.\ \ref{fig:LinesCheese}, 
outgoing from two two-dimensional slices of the superfluid which intersect 
the vortex cores along respectively perpendicular to the lines.
%
%
The vortices punch holes through the scalar charge density 
thus reducing the `screening' of the superfluid from the black hole.
The strongly coupled superfluid has been conjectured to exhibit an enhanced dissipation 
geometrically realized in holography
by modes falling through these holes into the black hole \cite{Adams:2012pj,Yang:2022foe}. 
In the two-dimensional superfluid, the holes are point-like and hence favor dissipation of UV modes. 
In the three-dimensional case, the holes are point-like in the directions perpendicular to a 
vortex line but spatially extended along the line, apparently permitting dissipation of large-wavelength modes of this direction. 
However, close to the vortex core the superfluid's density is small and its motion predominantly circular around
the vortex line. IR modes that could fall through the extended holes are thus absent at the
location of the holes, such that this dissipation mechanism favors UV modes also in three dimensions. 

\textit{Numerical simulation}. 
We study the superfluid's evolution starting from far-from-equilibrium initial 
conditions (see appendix \ref{app:IC})
and propagate the system in time until the last vortex defect has disappeared.
For that we solve the system's evolution on a periodic grid 
of $128^3$ points in the $\bm{x}$-directions and a basis of $32$ Chebyshev 
polynomials along the $z$-direction using
a fourth-order Runge--Kutta fixed-timestep algorithm.
In our results times are given as multiples of a unit timestep which is
composed of 250 numerical timesteps. 
The numerical implementation is described in appendix \ref{app:Num}.

\textit{Initial conditions}.
We prepare 24 straight vortex lines, composed of twelve pairs of 
opposite circulation with winding numbers $|w_i|=1$.
Of these, we randomly distribute four along each spatial direction $(x_1,x_2,x_3)$. 
In order to test the universality of the turbulent behavior we have also studied
the evolution with other initial conditions, including initial ensembles with vortex rings. 
Details of these are given in appendices \ref{app:IC} and \ref{app:TypeBDynamics}.

\begin{figure}
	\includegraphics[width=0.95\columnwidth]{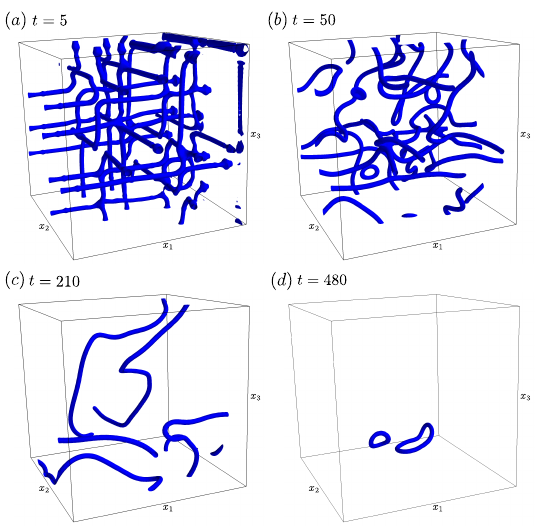}
	\caption{%
		Snapshots of isosurfaces of the superfluid density 
		$n(t,\bm{x})$ (defining value $n/n_0 = 0.23$) during  
		characteristic stages of the system's evolution: 
		$(a)$ The initial straight vortex lines are still discernible.
		$(b)$ Lines have reconnected several times and single rings have shrunk to zero size. 
		$(c)$ The vortex gas has significantly diluted and only few large vortex rings remain. 
		$(d)$ Only few small rings remain which quickly shrink to zero size.  
	\label{fig:Snapshots}
	}
\end{figure}

\textit{Evolution.} 
Figure \ref{fig:Snapshots} shows snapshots of the vortex configurations at four times characteristic for the system's 
evolution.{}\footnote{Videos showing the evolution of vortex lines and spectra as well as additional material can be found at 
\href{https://www.thphys.uni-heidelberg.de/~holography/Turbulence3D/}{https://www.thphys.uni-heidelberg.de/$\sim$holography/Turbulence3D/}.} 
Vortices are visible here as isosurfaces of the superfluid density $n(t,\bm{x})$. 
%
%
After the vortex lines have been imprinted, their flow fields  
and self-induced velocities immediately cause them to move and deform.\footnote{Flow patterns
  of simple vortex configurations like straight lines
  and rings in the holographic superfluid are illustrated in \cite{Ewerz:rings}.}
When cores of vortex line segments come into contact they reconnect, which can be 
resolved on a microscopic level in our simulation, 
see appendix \ref{app:Reconnection} for a more detailed discussion.
Within a short time, the system evolves from the initial ensemble of straight vortex lines into a
dense vortex tangle resembling those typically resulting from thermal quenches of superfluids. 
Repeated reconnections of the vortex lines lead to frequent formation of small vortex rings
which disappear by shrinking to zero size. In this process, the vortex gas dilutes to a state with
only few large rings. These often have elongated shape, such that further reconnections
can involve segments of the same ring. This typically leads to small, quickly contracting rings. Thermal
equilibrium is reached after all vortex lines have disappeared. 
The evolution proceeds in a similar way for other initial conditions,
see appendix \ref{app:TypeBDynamics}.
The rapid shrinking of small vortex rings provides evidence that the vortex dynamics is strongly dissipative. 
Holographic bulk views at different typical stages of the evolution are presented in appendix \ref{app:bulkhcheese}.

%
\textit{Turbulence} is characterized by quasi-stationary, algebraically decaying correlation functions. 
We consider the radial occupation number spectrum (or single-particle momentum 
spectrum) $n(t,k)$, given by the angle-averaged two-point correlator\footnote{%
  In practice, we follow \cite{Adams:2012pj,Ewerz:2014tua}
  and approximate the two-point 
	function $\langle\Psi\mbox{*}(t,\bm{k})\Psi(t,\bm{k})\rangle$ by the square 
	of the vacuum expectation value, $\vev{\Psi^\ast(t,\bm{k})}\vev{\Psi(t,\bm{k})}$, 
	as for the infrared physics of interest here quantum fluctuations are expected to be small
        compared to the vacuum expectation value.
}
\begin{equation}
       n(t,k)=\int\frac{\text{d}\Omega_k}{4\pi} \, \langle \Psi^\ast(t,\bm{k})\,\Psi(t,\bm{k}) \rangle\,,
\end{equation}
where $\bm{k}$ is the radial momentum vector Fourier-conjugate to $\bm{x}-\bm{y}$ with
an arbitrarily chosen reference point $\bm{y}$, and $k=|\bm{k}|$.
The kinetic energy spectrum is $E(k)=k^{4}n(k)$, valid for a dominating incompressible part \cite{Nore1997a}. 

The typical signature of turbulence is a quasi-stationary scaling 
$n(k)\sim k^{-\zeta}$ with scaling exponent $\zeta$ in a so-called inertial range of momenta.
Our procedure for determining the exponents and inertial ranges from the spectra is described in appendix \ref{app:Num}.

\begin{figure}[t]
  \includegraphics[width=0.9\columnwidth]{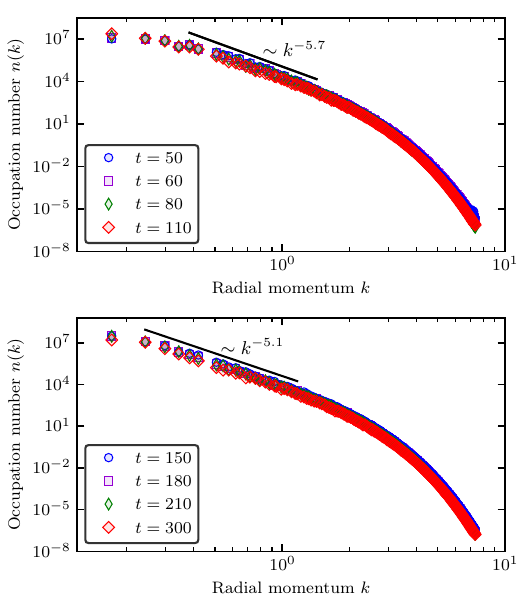}
	\caption{%
		Radial occupation number spectrum at different times during the 
		first (upper panel) and second (lower panel) universal scaling regime on double-logarithmic scales.
		The solid black lines indicate the fitted power-laws in the respective inertial momentum ranges. 
	\label{fig:SpecLines}
	}
\end{figure}

After the short initial stage of the evolution during which the vortex lines depart 
from their initial configuration and form a dense tangle,
the system enters a first quasi-stationary regime at approximately $t=50$. 
We observe a scaling behavior of the occupation number with exponent 
$\zeta=5.7\pm 0.2$ in the inertial momentum range $0.34 \le k \le 1.43$. 
This regime persists for $\Delta t \approx 60$ unit timesteps.
We show the spectrum for four times during this regime
in the upper panel of Fig.\ \ref{fig:SpecLines} where
the power law  $n(k)\sim k^{-5.7}$ is indicated as a solid black line. 
%
%
Our finding translates into $E(k)=k^{4}n(k)\sim k^{-1.7}$ for the kinetic energy spectrum. 
Within the uncertainty of our result, this resembles the power law $E(k) \sim k^{-5/3}$
characteristic for Kolmogorov scaling of classical turbulence \cite{Kolmogorov1941a,Kolmogorov1941b,Kolmogorov1941c}.
The vortex configuration for one exemplary time of this regime is displayed in panel $(b)$ of Fig.\ \ref{fig:Snapshots}. 

The system then transitions to a second quasi-stationary
scaling regime with exponent $\zeta=5.1\pm 0.2$ 
and inertial momentum range $0.24 \le k \le 1.12$.
This regime starts at approximately $t=150$ and persists for
about $\Delta t = 150$ unit timesteps, until at late times in the evolution 
only very few well-separated vortex rings remain. 
Snapshots of the spectrum at four exemplary times of this regime are
shown in the lower panel of Fig.\ \ref{fig:SpecLines}, and a 
characteristic vortex configuration is displayed 
in panel $(c)$ of Fig.\ \ref{fig:Snapshots}. 
The vortex gas has strongly diluted as compared to the prior 
scaling regime and the vortex lines reconnect only rarely. 

\begin{figure}[t]
	\includegraphics[width=0.9\columnwidth]{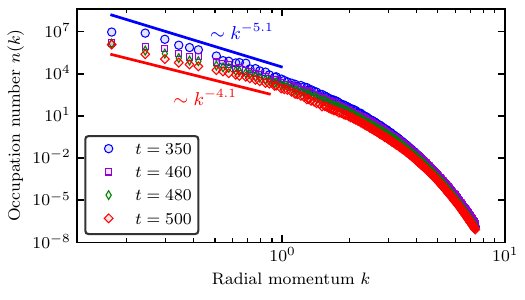}
	\caption{%
		Radial occupation number spectrum at four times during the late regime with only few rings remaining. 
                Here the spectrum continuously flattens 
                from the behavior $k^{-5.1}$ in the previous scaling regime to a power $k^{-4.1}$ shortly before the
                last ring disappears, as indicated by solid lines. 
	\label{fig:LinesEnd}
	}
\end{figure}
%
Once the vortex reconnections cease and only small rings are left, 
the spectrum deviates from the $n(k)\sim k^{-5.1}$ scaling behavior
and gradually flattens as the rings traverse the superfluid and shrink even further.
Fig.\ \ref{fig:LinesEnd} shows the spectrum at four 
exemplary times during this process, where $t=480$ corresponds to the 
vortex configuration in panel $(d)$ of Fig.\ \ref{fig:Snapshots}.
At $t=500$, just before the last ring vanishes, the spectrum is in the infrared
well approximated by $n(k)\sim k^{-4.1}$. 

We observe the same sequence of quasi-stationary scaling regimes and the
subsequent flattening of the spectrum also for other initial far-from-equilibrium
conditions of the system, \cf\ appendix \ref{app:TypeBDynamics}.
In all cases that we have studied, identical scaling exponents and inertial ranges
for the two scaling regimes are found within the accuracy of our numerical method. 
Only the times at which the system enters and leaves the scaling regimes 
depend to some extent on the initial condition. 
Hence, the occurrence of the two observed scaling regimes is a universal
property of the far-from-equilibrium dynamics of the three-dimensional holographic superfluid.

It is worth noting that turbulent scaling regimes with exponents close to the ones found here for the
holographic superfluid have been observed also in other models, including dilute superfluids,
see for example \cite{Nowak:2011sk}. 
Further study is required to determine whether these scalings originate from similar microscopic
dynamics in dilute systems and in the strongly coupled holographic superfluid. 

In the non-equilibrium dynamics of the two-dimensional 
holographic superfluid, a similar sequence of universal regimes 
of the occupation number spectrum has been found for various 
types of initial vortex configurations. 
In particular, a Kolmogorov-like scaling regime has been observed \cite{Adams:2012pj}
which is transient and is followed by a late-time $n(k)\sim k^{-4.1}$ scaling
regime \cite{Ewerz:2014tua}.
However, the late-time regimes in two and three dimensions are qualitatively different
due to topology. 
In the two-dimensional holographic superfluid the vortex dynamics at late times exhibits a critical
slowing down associated with algebraically decaying length scales \cite{Ewerz:2014tua}, 
which has been interpreted as the approach to a
non-thermal fixed point \cite{Berges:2008wm,Berges:2008sr,Scheppach:2009wu} of the evolution.
We do not find evidence for an analogous critical slowing down in three dimensions.
In two dimensions a dense ensemble of vortices and anti-vortices generically evolves to a
system of few widely separated and thus slowly moving vortex defects, and thermal equilibrium
is only reached after the pairwise annihilation of all defects. 
In three dimensions, on the other hand, we observe that a dense vortex tangle generically evolves
into a set of small vortex rings which rapidly shrink and annihilate separately.
This process is accelerated by splittings of elongated rings into smaller rings via reconnections.
However, configurations of large and widely separated vortex rings, as might correspond to a non-thermal
fixed point, are in principle possible. But it would require distinctly larger numerical
lattices than we use here to observe them and to study whether they can occur generically at late times.

\textit{Summary.} 
Real-time simulations of far-from-equilibrium evolution in a three-dimensional
holographic superfluid exhibit two subsequent universal regimes of quasi-stationary scaling,
characterized by the power-laws $k^{-5.7}$ and $k^{-5.1}$ of the occupation number spectrum,
the former being Kolomogorov-like. The quantum vortex dynamics is found to be strongly dissipative. 
The methods developed here will permit to study a wide range of phenomena in strongly dissipative
superfluids in three dimensions, promising new insights into vortex dynamics in real-world superfluids. 

\textit{Note added.} 
While this paper was in preparation, we received \cite{Zeng:2024rwn} which has some overlap with the present work.

\textit{Acknowledgments.} 
We thank G.\ Bals, T.\ Gasenzer, D.\ Proment, A.\ Samberg and C.--M.\ Schmied for helpful discussions.  
P.\,W.\ was supported by the Studienstiftung des deutschen Volkes e.V. 

\bibliographystyle{apsrev4-1}

\clearpage
\begin{appendix}
	\begin{center}
		\onecolumngrid
		\textbf{\large Appendix}
	\end{center}
	\setcounter{equation}{0}
	\setcounter{table}{0}
	\makeatletter
		
\section{Holographic superfluid}
\label{app:HoloSetup}
%
Here we review the gravitational model of the three-dimensional superfluid studied in the main text.
The action \eqref{eq:ActionAdSCFT} describes the simplest holographic model of
a superfluid. Note that symmetry breaking and thus superfluidity 
can occur in this model already without a $|\Phi|^4$ term \cite{Gubser:2008px}. 
However, such a term or a more general potential for $\Phi$ could be added, giving rise to a class of different
holographic models of superfluids. 
We emphasize that the model considered here is a bottom-up holographic model. 
While the gravity side of a bottom-up model is well known by construction, the Lagrangian of the dual field theory is not explicitly known.
As the bottom-up construction assumes a weakly coupled gravity side, the dual field theory can be inferred 
to be strongly coupled since holographic dualities are generically of strong-weak type.
However, the value of the coupling and of other phenomenological parameters of the field theory is not a priori known
in a bottom-up model. 

In the following, we provide the explicit equations of motion and the boundary conditions
of the gauge--matter fields for the model \eqref{eq:ActionAdSCFT} of the three-dimensional holographic superfluid. 
We furthermore lay out how one extracts the 
superfluid order-parameter field from solutions of the gauge--matter fields.
In this work, natural units are used in which $\hbar, c, k_\text{B}=1$. 
We work in the probe approximation in which the 
backreaction of the gauge--matter sector on the gravitational sector is neglected. 
This approximation is applicable if the temperature of the system is 
below but of the order of the phase-transition temperature.
Formally, large $q$ corresponds to the same probe limit, where $q$ is the charge of the scalar field which effectively 
controls the coupling of the gauge--matter fields to the gravitational 
background (\cf\ the action \eqref{eq:ActionAdSCFT}), see \cite{Herzog:2008he, Albash:2009iq, Sonner:2010yx}.
Using this approximation, the gravitational part of the 
action \eqref{eq:ActionAdSCFT} reduces to the Einstein--Hilbert action with a negative cosmological constant. 
It is solved by a $(4+1)$-dimensional anti-de Sitter spacetime with a 
planar Schwarzschild black hole.
The line element is
\begin{equation}\label{eq:BackMetric}
	\td s^2  =  \frac{L_\text{AdS}^2}{z^2}\left(-h(z)\,\td t^2+\td\bm{x}^2-2\td t\,\td z\right)\,,
\end{equation}
here written in infalling Eddington--Finkelstein coordinates with respect to the holographic coordinate $z$. 
$L_\text{AdS}$ is the curvature radius of the AdS$_5$ spacetime 
which in the following we set to unity, $L_\text{AdS}=1$, for convenience.

As in the main text, we use $\bm{x}=(x_1,x_2,x_3)$ to denote 
the three spatial field-theory directions. 
The horizon function is given by
\begin{equation}
	h(z)=1-\left(\frac{z}{\zh}\right)^4\,,
\end{equation}
where $\zh$ is the position of the black-hole horizon along the holographic $z$-direction. 
The temperature of the superfluid coincides with the Hawking temperature of the black hole
\begin{equation}
	T=\frac{1}{\pi\zh}\,.
\end{equation}
As we neglect the backreaction of the gauge--matter sector on the 
black hole, the temperature does not fluctuate and remains constant in $\bm{x}$ and $t$. 

Given the constant gravitational background, one can derive 
the equations of motion for the gauge--matter fields $A_\mu$ and $\Phi$. 
They are given by the Klein--Gordon and Maxwell equations
\begin{align}
\left( D^2 + m^2 \right) \Phi &= 0\label{eq:KGE}\,,\\
  \nabla_{\mu}F^{\mu \nu} &=J^\nu =  \text{i}  \left[\Phi^* D^{\nu} \Phi - \Phi \left ( D^{\nu} \Phi\right)^*\right]
                            \,,\label{eq:Maxwell}
\end{align}
with the gauge covariant derivative $D_{\mu}=\nabla_\mu - \text{i} A_\mu$, 
where $\nabla_\mu$ denotes the covariant derivative associated with the metric \eqref{eq:BackMetric}.
Note that we have rescaled the fields $\Phi, A_\mu$ such that the charge $q$ of the scalar controls the coupling of the gauge--matter 
sector $\mathcal{L}_{\mathrm{gm}}$ to the gravitational sector.

For a static and spatially homogeneous system 
in $\bm{x}$, the fields only depend
on the holographic $z$-direction. 
We first solve the system of equations of motion \eqref{eq:KGE}--\eqref{eq:Maxwell} 
with this simplifying assumption to find the constant background density of the superfluid,
see \cite{Horowitz:2008bn,Arean:2010zw} for similar considerations. 
For the gauge field, the Maxwell equations reduce to
\begin{align}
0&=z^2\,A_t''-z\,A_t'+2\,\text{Im}\,({\Phi}'\,{\Phi}\mbox{*})\label{eq:AtStatic}\,,\\
0&= z^2\,\left(h\,A_i''+A_i'\,h'\right)-z\,h\,A_i'-2\,A_i\,|\Phi|^2\label{eq:AiStatic}\,,\\
0&=2\,A_z\,|\Phi|^2+2\,h\,\text{Im}({\Phi}'\,{\Phi}\mbox{*})\label{eq:AzStatic}\,,
\end{align}
where we use $i=(x_1,x_2,x_3)$ for the three spatial boundary 
directions and denote derivatives with respect to the holographic $z$-direction by primes. 
In addition, for the scalar field one finds
\begin{align}\label{eq:ScalarFieldStatic}
0=z^2\,h\,\Phi''  -z\left(-2\text{i}z\,A_t+3h-z\,h'\right)\Phi'  -  \left(m^2+z^2\,\bm{A}^2+3\text{i}z\,A_t-\text{i}z^2\,A_t'\right)\Phi\,,
\end{align}
where we introduced $\bm{A}=(A_{x_1},A_{x_2},A_{x_3})$.
We fix the gauge freedom by setting $A_z=0$. 
Equation \eqref{eq:AzStatic} is therefore used as a constraint to ensure the chosen gauge condition.  
The boundary conditions for the fields are as follows: 
\begin{align}\label{eq:BoundaryConditions}
A_t (z = 0) &= \mu\,, \, &A_t(z=\zh) = 0\,, \nonumber\\
A_{x_1} (z = 0) &= 0\,, \, &A_{x_1}(z=\zh) = 0\,, \nonumber\\
A_{x_2} (z = 0) &= 0\,, \, &A_{x_2}(z=\zh) = 0\,, \\
A_{x_3} (z = 0) &= 0\,, \, &A_{x_3}(z=\zh) = 0\,,\nonumber \\
\partial_z \Phi (z) \, \lvert_{z=0} &= 0\,, & \vert \Phi(z=\zh)\vert<\infty \,,   \nonumber
\end{align}
where the last of these is a behavioral boundary condition for the scalar 
field $\Phi$, implying regularity at the black-hole horizon $z=\zh$.
The boundary value of the temporal gauge-field component is given by the chemical potential $\mu$.

\begin{figure}
	\includegraphics[scale=0.8]{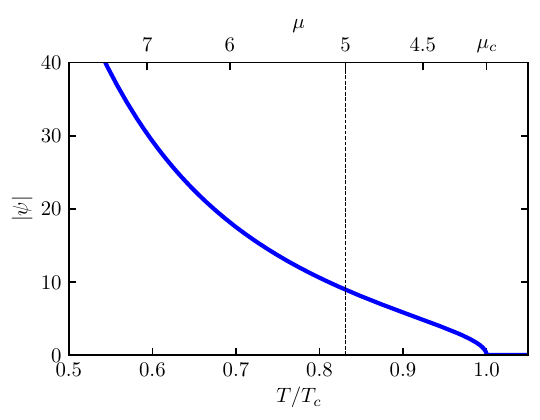}
	\caption{%
          Modulus $|\psi|$ of the superfluid order parameter for the static background solution 
          as a function of temperature $T$ 
		(lower abscissa) or chemical potential $\mu$ (upper abscissa).
		Below the critical temperature $\Tc$ (or above the critical chemical potential $\mu_c$) 
		the order parameter is non-zero and the system is in the superfluid phase. 
		The dashed black line indicates the temperature resp.\ chemical potential used in this work. 
	\label{fig:SuperCond}
	}
\end{figure}

Solving the set of coupled equations of motion \eqref{eq:AtStatic}--\eqref{eq:ScalarFieldStatic} 
with the appropriate boundary conditions \eqref{eq:BoundaryConditions} 
gives the background solution for the gauge--matter fields. 
The mass of the scalar field can be chosen freely, as long as its square is negative and above 
the Breitenlohner--Freedman bound \cite{Breitenlohner:1982bm,Breitenlohner:1982jf}. 
We choose $m^2=-3$ which is convenient with regard to the extraction of the superfluid order parameter from the 
near-boundary expansion of the scalar field $\Phi$.
For the given mass-squared, the near-boundary expansion reads
\begin{equation}\label{eq:PhiBoundaryExp}
        \Phi(t, \bm{x}, z)=\eta(t, \bm{x})\,z +  \psi(t,\bm{x}) \,z^3 +\mathcal{O}(z^4)\,,
\end{equation}
where $\eta$ is the source conjugate to the dual scalar 
operator $\Psi$, and $\psi=\vev{\Psi}$ is its expectation value.
$\Psi$ is a relevant operator if $m^2<0$. 
The boundary condition for $\Phi$ at $z=0$ in \eqref{eq:BoundaryConditions} is chosen such that 
the source is set to zero which ensures that the $U(1)$ symmetry is not broken explicitly.
The complex superfluid order-parameter field $ \psi(t,\bm{x})$ can then 
straightforwardly be extracted from \eqref{eq:PhiBoundaryExp}.
The phase transition is controlled by the single dimensionless 
parameter $\mu/T$ \cite{Gubser:2008px,Hartnoll:2008vx}. 
To fix our units we set $\zh=1$ for the present work which leaves 
the chemical potential $\mu$ as the only free parameter. 
Varying $\mu$ corresponds to varying the critical temperature of the phase transition due
to the relation $\mu T/\Tc = \mu_c$. 
Above the critical chemical potential $\mu_c \simeq 4.1568$, the system 
is in the superfluid phase. The background density $n_0$ is obtained from the static
background solution by $n_0=| \psi|^2$, and we have $n_0> 0$ in the superfluid phase. 
In Fig.\ \ref{fig:SuperCond} we show the modulus of the 
superfluid order parameter $\psi$ for the static solution as a function of the 
temperature ratio $T/\Tc$ or of the chemical potential $\mu$.  
%
%
The dashed black line indicates the chemical potential used in this 
work, $\mu=5$, which sets the system in the superfluid phase at $T/\Tc=0.83$.

In our numerical implementation we first solve the static equations of motion 
discussed above and then imprint vortex defects onto the superfluid 
density (see appendix \ref{app:IC} for details). 
To propagate the system in time we have to solve the full set of 
equations of motion \eqref{eq:KGE}--\eqref{eq:Maxwell}, without 
further simplifying assumptions. 
In view of the near-boundary expansion of the scalar field \eqref{eq:PhiBoundaryExp}, it is useful to define the rescaled field
\begin{align}\label{eq:AtidleEq}
\tilde{\Phi}\equiv \frac{\Phi}{z} \,,
\end{align}
in terms of which we formulate the equations of motion.
As for the static equations we use the gauge $A_z=0$. 
Using the metric \eqref{eq:BackMetric} to define the gravitational covariant derivative, the equations 
of motion \eqref{eq:KGE}--\eqref{eq:Maxwell} for the spatio-temporal components of the gauge 
field and for the scalar field, after some algebra, reduce to
\begin{align}
	\partial_zA_t-z\partial^2_zA_t~=~
	&2z\,\text{Im}(\tilde{\Phi}\mbox{*}\partial_z\tilde{\Phi})-z\partial_z\bm{\nabla}\cdot\bm{A} \,,\label{eq:A0PDE}\\[.6em]
	\partial_t(A_{x_1}-2z\partial_zA_{x_1})~=~
	&z\,\Big[\partial_{x_1}\big(\partial_{x_2}A_{x_2}+\partial_{x_3}A_{x_3}-\partial_zA_t\big)-(\partial_{x_2}^2+\partial_{x_3}^2)A_{x_1}\Big]
	+\partial_{x_1}A_t-2z\,\text{Im}(\tilde{\Phi}\mbox{*}\partial_{x_1}\tilde{\Phi})\label{eq:Ax1PDE}\nonumber\\
	&+\big(1+3z^4\big)\partial_zA_{x_1}-hz\partial^2_zA_{x_1}+2z|\tilde{\Phi}|^2A_{x_1}\,,\\[.6em]
	\partial_t(A_{x_2}-2z\partial_zA_{x_2})~=~
	&z\,\Big[\partial_{x_2}\big(\partial_{x_1}A_{x_1}+\partial_{x_3}A_{x_3}-\partial_zA_t\big)-(\partial_{x_1}^2+\partial_{x_3}^2)A_{x_2}\Big]
	+\partial_{x_2}A_t - 2z\,\text{Im}(\tilde{\Phi}\mbox{*}\partial_{x_2}\tilde{\Phi})\label{eq:Ax2PDE}\nonumber\\
	&+\big(1+3z^4\big)\partial_zA_{x_2}-hz\partial^2_zA_{x_2}+2z|\tilde{\Phi}|^2A_{x_2}\,,\\[.6em]
	\partial_t(A_{x_3}-2z\partial_zA_{x_3})~=~
	&z\,\Big[\partial_{x_3}\big(\partial_{x_1}A_{x_1}+\partial_{x_2}A_{x_2}-\partial_zA_t\big)-(\partial_{x_1}^2+\partial_{x_2}^2)A_{x_3}\Big]
	+\partial_{x_3}A_t-2z\,\text{Im}(\tilde{\Phi}\mbox{*}\partial_{x_3}\tilde{\Phi})\label{eq:Ax3PDE}\nonumber\\
	&+\big(1+3z^4\big)\partial_zA_{x_3}-hz\partial^2_zA_{x_3}+2z|\tilde{\Phi}|^2A_{x_3}\,,\\[.6em]
	\partial_t\big(\tilde{\Phi}-2z\partial_z\tilde{\Phi}\big)~=~
	&z \left[\text{i}\big(\bm{\nabla}\cdot\bm{A}-\partial_zA_t\big)\tilde{\Phi}+
	2\text{i}\bm{A}\cdot\bm{\nabla}\tilde{\Phi}-h\partial_z^2\tilde{\Phi}-\mathbf{\bm{\nabla}}^2\tilde{\Phi} \right]\nonumber\\
	&+\text{i}A_t\tilde{\Phi}-2\text{i}zA_t\partial_z\tilde{\Phi}+(z^3+z\bm{A}^2)\tilde{\Phi}
	+(1+3z^4)\partial_z\,\tilde{\Phi}\,,\label{eq:PhiPDE}
\end{align}
where we have suppressed all functional dependencies to aid the readability. 
We recall that the gauge field $A_\mu$ as well as the scalar field $\tilde\Phi$ depend on all five 
coordinates, $(t,\bm{x},z)$, while the horizon function $h=h(z)$ depends 
only on the holographic $z$-coordinate. 
In the above equations we have used the gradient $\bm{\nabla}=(\partial_{x_1},\partial_{x_2},\partial_{x_3})$ 
with respect to the three spatial field-theory directions. 

Finally, we find from the remaining $z$-component of the Maxwell equations \eqref{eq:Maxwell}
\begin{align}\label{eq:AzDynamicConstraint}
	0 ~=~ \del_t\del_zA_t - \bm{\nabla}^2A_t + \del_t\bm{\nabla}\cdot\bm{A}
	- h\del_z\bm{\nabla}\cdot\bm{A} + 2\lvert\Phit\rvert^2A_t 
		- 2\,\text{Im}\left(\Phit^*\del_t\Phit - h\Phit^*\del_z\Phit\right) \,.
\end{align}
Like for the static system discussed in the previous section, this equation is 
not independent of equations \eqref{eq:A0PDE}--\eqref{eq:PhiPDE}.
Instead, it can be expressed as a derivative of a combination of equations \eqref{eq:A0PDE}--\eqref{eq:PhiPDE}.
Specifically, it can be shown that if equations \eqref{eq:A0PDE}--\eqref{eq:PhiPDE} are 
satisfied, equation \eqref{eq:AzDynamicConstraint} is satisfied for all $z$ if it is satisfied on one fixed-$z$ slice in the bulk. 

Equations \eqref{eq:A0PDE}--\eqref{eq:PhiPDE} are a closed 
system of partial nonlinear differential 
equations in the five coordinates $t,\bm{x}$ and $z$.
When solving them we 
use the same boundary conditions for all $(t,x_1,x_2,x_3)$ as for the static case, \cf\ \eqref{eq:BoundaryConditions}.

\section{Initial conditions}\label{app:IC}
%
In this appendix we discuss the initialization of ensembles of vortex lines in the superfluid which we use as initial
conditions for our simulations. Within a few timesteps, these initial vortex-line configurations develop into
a tangle of vortex lines akin to those typically expected to emerge from a rapid quench of the system,
see \cite{Nowak:2012gd} for a discussion of quenches in GP dynamics. The emergence of vortex defects from 
a quench into the superfluid phase has been observed experimentally for example in \cite{Hendry1994}.

Straight vortex lines can be constructed based on  
the well-known procedure for imprinting a vortex--anti-vortex pair in a holographic superfluid 
in two spatial dimensions \cite{Keranen:2009re} (see also \cite{Ewerz:2020wyp}), making use of 
translational symmetry along the additional direction. We consider the example of a straight vortex 
line oriented along the $x_3$-axis. To initialize such a vortex defect we 
imprint a two-dimensional vortex onto the background density
at the same $(x_1, x_2)$-position of every $x_3$-slice. 
This is done by multiplying into the static solution of the complex scalar field $\Phi(\bm{x},z)$ for 
every $z$ (the holographic direction) a phase $\varphi(\bm{x})$ 
which winds around the position $\bm{x}_0$ on all closed loops in the $(x_1, x_2)$-plane
encircling $\bm{x}_0$ once by $2\pi$ or $-2\pi$ (or multiples thereof if one were to
initialize vortex lines of higher winding number).
It is convenient (but not strictly necessary) to choose the phase $\varphi(\bm{x})$ linear in the
geometric angle of a circle enclosing $\bm{x}_0$. 
In addition, we set the superfluid density to zero at the position of the vortex core. 
We point out that we do the latter only to accelerate the dynamical build-up process of the vortex solutions.
Imprinting only the phase winding would suffice as the superfluid density would drop to zero within a few
timesteps of the numerical evolution due to the topological constraint. 
By symmetry, the analogous procedure can be applied for any  
differently oriented straight vortex line. 
For an ensemble of $N$ vortex lines the phases are added up 
linearly $\varphi(\bm{x})=\sum_{i=1}^{N}\varphi_i(\bm{x})$ and 
then multiplied into the static solution of the complex scalar field, $\Phi(\bm{x},z) \to \Phi(\bm{x},z)\,e^{\i\varphi(\bm{x})}$.
It should be noted that the periodicity of the computational domain in $(x_1, x_2,x_3)$
requires that the total winding number of all straight vortex lines vanishes along each coordinate direction. 
The configurations thus constructed do not immediately correspond to physical vortex lines. 
But when evolving the configurations in time they quickly approach physical vortex lines with
their typical density profile. It takes about 5 unit timesteps until the vortex profiles are fully developed and stable.
Also the phase field quickly approaches a physical configuration in this process. However, small deviations of the phase field
from an exact physical configuration take significantly longer to disappear during the numerical evolution. An analogous phase-healing 
has been observed and quantified in \cite{Ewerz:2020wyp} for vortices in the two-dimensional superfluid.
We emphasize that the corresponding deviations
in the vortex-line trajectories caused by these phase-field deviations have a size significantly below the distance between
neighboring grid points of the numerical lattice. They do not affect the occupation number spectra discussed here in a discernible way. 

Analogously, using azimuthal symmetry instead of the translational symmetry, closed vortex rings can be imprinted.
A more detailed exposition of closed vortex rings and their dynamics in the holographic superfluid
will be given elsewhere \cite{Ewerz:rings}. For a discussion of vortex ring dynamics in the Gross--Pitaevskii framework
see for example \cite{Berloff2007}. 

For the initial conditions of the simulations presented in the main text, in the following referred to
as initial condition $\mathcal{A}$, we prepare a total of 24 vortex lines, divided 
into twelve pairs of anti-parallel defects, \ie, the vortex lines of each pair have opposite 
circulation, with winding numbers \mbox{$|w_i|=1$}.
Of these twelve pairs of vortex--anti-vortex lines, we randomly distribute 
four along each of the three spatial directions $(x_1,x_2,x_3)$.
We ensure that initially 
the vortex cores do not intersect at any point.
To distribute them randomly we use a Gaussian random distribution. 

In order to test the universality of the turbulent behavior of the system, we performed
various simulations starting from different initial conditions. 
Specifically, in addition to the initial condition of type $\mathcal{A}$ consisting of vortex lines (see above),
we also prepare another type given by a configuration of initial vortex rings. 
This initial condition, in the following called initial condition $\mathcal{B}$, is 
given by $18$ plane circular vortex rings.
We randomly distribute the centers of these vortex 
rings over the entire three-dimensional 
domain and align their symmetry axes along the three 
coordinate axes of the grid, six along each. 
Of these six vortex rings, three have winding number $w_i=+1$ and 
the other three have winding number $w_i=-1$.
Like for the vortex positions, we also employ a 
Gaussian random distribution to set the initial radii $R_i$ of the rings. 
However, we impose the constraint $5<R_i<64$ (in grid points) which ensures 
that the maximal diameter of every individual 
ring does not exceed the grid size along any of the three coordinate directions. 
We further impose as a constraint for the vortex rings in the initial configuration 
that no intersections of any two vortex cores occur 
at any point in the computational domain (to inter-mesh accuracy). 
Intertwined vortex rings, on the other hand, are permitted.

To verify that the scaling behavior discussed in the main text is 
indeed universal, we also prepared various types of initial conditions of 
type  $\mathcal{A}$ and $\mathcal{B}$ with varying numbers of initially imprinted 
vortex lines or rings as well as mixtures of type $\mathcal{A}$ and $\mathcal{B}$ (straight lines plus rings)
with different numbers of defects. The simulations with combinations of lines and rings in the initial condition
were performed on smaller grids of $64^3$ and $96^3$ points. 

\section{Numerical implementation}\label{app:Num}
%
In the following, we discuss the details of the numerical methods used to 
solve the differential equations presented in appendix \ref{app:HoloSetup}.
Furthermore, we discuss the power-law fitting of the occupation number spectra  
and how we estimate the uncertainty of the corresponding scaling exponents.

\subsection{Numerical implementation of the equations of motion}
%
By construction, the equations of motion for the static background 
solution of the superfluid depend only on the holographic $z$-coordinate  
on a computational domain set by the Schwarzschild--AdS spacetime, $0\leq z \leq \zh$.
We expand the fields in a basis of $32$ Chebyshev polynomials on a 
Gauss--Lobatto grid which allows us to implement derivatives via matrix multiplication. 
As the equations are nonlinear, we solve them iteratively using a 
Newton--Kantorovich procedure which linearizes them. 
At each step we then solve a set of linear equations. 

For the time evolution of the full system
we further expand the fields in a basis of $128$ Fourier modes  
along each of the $(x_1,x_2,x_3)$-directions, choosing a grid spacing 
of $a=1/3.5$ (in units of $\zh=1$).
This allows us to implement derivatives with respect to the 
spatial field-theory directions $(x_1,x_2,x_3)$ via discrete Fourier transforms. 
For the $z$-direction we again proceed as outlined above. 

Equations \eqref{eq:A0PDE}--\eqref{eq:PhiPDE} are a closed 
system of partial nonlinear differential 
equations in the five coordinates $t,\bm{x}$ and $z$.
While the equations appear rather challenging to solve at first sight, their structure 
allows an essentially straightforward numerical implementation.
We first note that derivatives with respect to time only 
appear on the left-hand sides of equations \eqref{eq:Ax1PDE}--\eqref{eq:PhiPDE}. 
On a fixed timeslice, we can therefore compute the right-hand sides at 
every point $(x_1,x_2,x_3,z)$ and then integrate the respective outcomes with respect to time.
Subsequently, we solve the remaining linear differential 
equations in $z$ to obtain the fields $\bm{A}$ and $\tilde{\Phi}$ on the new timeslice. 
Finally, we use these fields to solve equation \eqref{eq:A0PDE} to obtain
the temporal gauge-field component on the new timeslice. 
For the boundary conditions for the fields, we 
again employ equations \eqref{eq:BoundaryConditions}.
We note that on the respective left-hand sides of the 
dynamical equations \eqref{eq:Ax1PDE}--\eqref{eq:Ax3PDE}, for the 
spatial gauge-field components $\bm{A}$, and \eqref{eq:PhiPDE}, for the scalar 
field $\tilde\Phi$, only first-order derivatives with respect to $z$ occur.
Consequently, in the intermediate step when we solve the linear differential 
equations in $z$ for the fields $\tilde\Phi$ and $\bm{A}$, we need 
to impose boundary conditions only at $z=0$.
For $A_t$, on the other hand, equation \eqref{eq:A0PDE} is of second order in $z$ and 
we therefore impose boundary conditions at $z=0$ and $z=\zh$.
We implement the boundary conditions such that $\mu = A_t(z=0)$ is strictly
conserved throughout the evolution, corresponding to the grand canonical ensemble. 

For the time propagation, we employ a fourth-order fixed-timestep Runge--Kutta algorithm. 
In units of $\zh=1$, we choose the timestep 
size such that one unit timestep is composed of $250$ numerical timesteps. 

The solving algorithm discussed above is implemented in \Cpp. 
For Fourier transforms we use the \texttt{fftw3} library \cite{Frigo:2005zln} and 
implement all linear-algebra operations with help of the \texttt{Eigen} library \cite{eigenweb}.
The code is parallelized with \texttt{OpenMP} \cite{OpenMP}.
We use \texttt{Mayavi} \cite{ramachandran2011mayavi} for all three-dimensional visualizations. 

\subsection{Extracting the scaling exponents of the occupation number spectra}\label{app:Scaling}
%
Scaling behavior in the radial occupation number spectrum 
is characterized by $n(k)\sim k^{-\zeta}$ within a certain inertial 
momentum range, where $\zeta$ is the scaling exponent.
We extract the scaling exponent from our data points using a  
Levenberg--Marquardt least-squares fitting routine, following the method used in
\cite{Nowak:2011sk,Ewerz:2014tua}. 

This procedure is in general plagued by a number of uncertainties.
First and foremost, the determination of the inertial momentum range. 
We initially choose the endpoints of the momentum interval by eye and 
subsequently vary them systematically to estimate 
the uncertainty of the scaling exponent. 
We find the corresponding uncertainty to be given by $\Delta \zeta=\pm0.2$. 
The fitting routine itself, on the other hand, typically reports an uncertainty 
of \mbox{$\Delta \zeta=\pm 0.05$} which is clearly subdominant.
Yet another source of uncertainty comes from the noise in the set of 
data points due to the finite size of the computational domain. 
This uncertainty is difficult to determine. To estimate it, we 
perform simulations similar to those presented in the main 
text with analogous initial conditions on smaller grids 
of $96^3$ points along the spatial field-theory directions 
and evaluate the sets of data points using the same methods as outlined above.
Comparing the results for the scaling exponents 
yields discrepancies of the order of $\Delta \zeta=\pm 0.1$,
which is slightly smaller than the uncertainty 
originating from the choice of endpoints of the inertial momentum range.
Naturally, we expect the finite-size-induced noise to decrease if we would work on still larger numerical grids or 
average the results over several simulation runs with similar initial vortex configurations. 
Presently, however, both of these methods appear out of reach given the available 
computational resources which strongly constrain the feasible grid 
sizes and number of simulation runs we can perform.
Finally, the different uncertainties discussed above are not independent 
as, for instance, noise in the data can affect the determination of the inertial momentum range. 
Considering all sources of uncertainty combined, we choose to 
quote $\Delta \zeta=\pm0.2$ as our best estimate for the uncertainty of the 
scaling exponents throughout this work.

\section{Scaling behavior for different initial conditions}\label{app:TypeBDynamics}
%
%
In this appendix we discuss the superfluid's scaling behavior starting from 
initial condition $\mathcal{B}$, consisting of $18$ vortex rings 
randomly distributed in the computational domain (for details see appendix \ref{app:IC}).\footnote{For videos for initial condition $\mathcal{B}$ see 
  \href{https://www.thphys.uni-heidelberg.de/~holography/Turbulence3D/}{https://www.thphys.uni-heidelberg.de/$\sim$holography/Turbulence3D/}. \hspace*{4cm}}

\begin{figure}[t]
	\centering
	\includegraphics[width=0.49\linewidth]{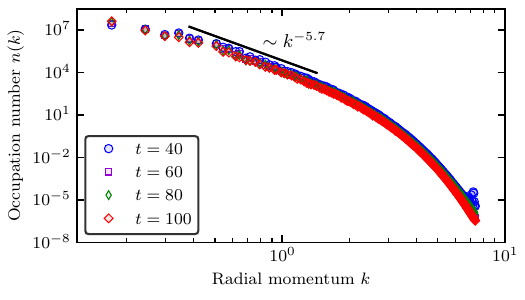}
	\includegraphics[width=0.49\linewidth]{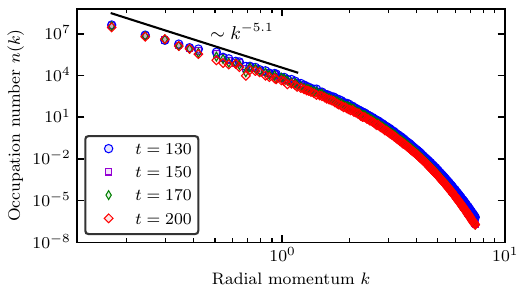}
	\caption{%
          Occupation number spectrum $n(k)$ for initial condition $\mathcal{B}$ (vortex rings)
          as a function of the radial momentum $k$ at 
		four characteristic times during the first (left panel) and 
		second (right panel) universal regime, on double-logarithmic scales. 
		The solid black lines indicate the fitted power laws in the respective inertial 
		momentum ranges (deliberately shifted above the spectra).
				The scaling exponents as well as the inertial momentum ranges for 
				which the spectrum shows scaling behavior agree with the results 
				for initial condition $\mathcal{A}$ (vortex lines) discussed in the main text. 
	\label{fig:TypeBSpectra}
	}
\end{figure}

For initial condition $\mathcal{B}$, it takes approximately $\Delta t=40$ 
unit timesteps until the first scaling regime of the occupation number spectrum is entered. 
At this point in the evolution, the system has largely lost 
memory of the details of its initial condition. 
The vortical excitations of the system consist of numerous 
closed vortex rings of different shapes and sizes. 
In addition, several small vortex rings have already disappeared by shrinking to zero size.
The first scaling regime has a scaling exponent of $\zeta=5.7\pm0.2$ in the inertial momentum range
$0.34 \le k \le 1.43$, \cf\ the left panel in Fig.\ \ref{fig:TypeBSpectra}. This agrees 
with the scaling law and inertial range we find for initial condition $\mathcal{A}$ (see main text).
We note that also here this Kolmogorov-like scaling regime is only transient. 
It persists for approximately $\Delta t= 60$ unit timesteps before the system transitions to the second scaling regime.
This second regime is entered at $t=130$ and persists for approximately $\Delta t= 70$ unit timesteps. 
We show spectra corresponding to the second scaling regime in the right panel of Fig.\ \ref{fig:TypeBSpectra}.
Again, the scaling exponent of $\zeta=5.1\pm0.2$ agrees with the result obtained for initial condition
$\mathcal{A}$ presented in the main text. The inertial momentum range is $0.23 \le k \le 1.12$, and is to the accuracy
of its determination in agreement with the one found for initial condition $\mathcal{A}$. 
We note that the duration of the respective scaling regimes depends on the initial conditions and is,
in contrast to the scaling exponents and the corresponding inertial momentum ranges, not universal.

\begin{figure}[t]
	\centering
	\includegraphics[width=0.49\linewidth]{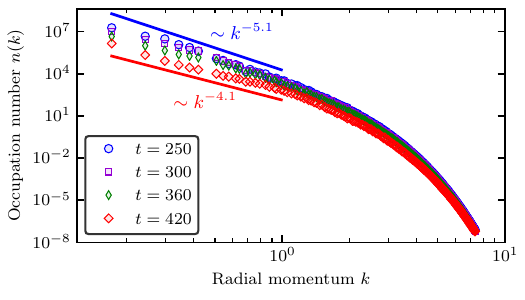}
	\caption{%
          Occupation number spectra $n(k)$ for initial condition $\mathcal{B}$ (vortex rings)
          at late times of the evolution when only few 
		non-reconnecting vortex rings remain and rapidly shrink to zero size. 
		As the rings contract, the spectrum gradually flattens. 
                The straight lines correspond to power-laws and indicate the evolution from 
                the scaling behavior in the prior scaling regime ($k^{-5.1}$) to the 
                approximate power law ($k^{-4.1}$) shortly before the final ring has disappeared.
		This behavior again agrees with the one observed for an initial 
		configuration of straight lines (initial condition $\mathcal{A}$) as discussed in the main text.
	\label{fig:TypeBSpectraLate}
	}
\end{figure}

Once there are only few well-separated vortex rings left, the spectrum 
deviates from the $\zeta=5.1$ scaling and gradually flattens.
We show the spectrum at four different late times in Fig.\ \ref{fig:TypeBSpectraLate}.
The behavior again agrees with the one for initial condition $\mathcal{A}$.

In addition, we analyzed the spectra for other initial conditions, see appendix \ref{app:IC}. 
Due to numerical constraints these simulations were 
performed on smaller grids of $64^3$ and $96^3$ points. 
We find that the spectra again agree with the results presented 
here and in the main text. 
In particular, we have checked explicitly that these results  
are independent of the number of initial vortex lines and vortex rings as long 
as the numbers are large enough to create a dense vortex gas. 
We therefore conjecture that the 
dynamics exhibits universal behavior, which persists until the final vortex ring has 
disappeared and the system has equilibrated, with two regimes in which the 
system exhibits (quasi-)stationary scaling with the scaling exponents and inertial momentum ranges
specified above.

It is worth noting that scaling regimes similar to those found in the holographic superfluid
have been observed already in the non-dissipative GP model, both in two and three dimensions
\cite{Nore1997a,Nore1997b,Araki2002,Kobayashi2002,Nowak:2010tm,Nowak:2011sk,Nowak:2012gd,Mathey:2014xxa}.
However, further study is needed to determine whether the scaling behavior has the same origin
in GP dynamics and in the strongly dissipative holographic superfluid. This would require
in particular a detailed analysis of the energy cascades in the respective regimes. 

\section{Vortex reconnections and shrinking rings}\label{app:Reconnection}
%
In this appendix we discuss how vortex lines and rings 
in the holographic superfluid reconnect and how vortex rings disappear by shrinking to zero size.\footnote{For a video of vortex reconnections see 
  \href{https://www.thphys.uni-heidelberg.de/~holography/Turbulence3D/}{https://www.thphys.uni-heidelberg.de/$\sim$holography/Turbulence3D/}. \hspace*{4cm}}
Illustrations of the flow field of simple vortex-line configurations are given in \cite{Ewerz:rings}. 

For the purpose of studying vortex reconnections, we use an initial condition consisting of straight vortex lines. 
Due to the periodicity of the computational $(x_1,x_2,x_3)$-domain 
the total winding number of straight vortex lines in the initial configuration along each spatial 
direction has to be zero. 
We first note that for the simplest case of two anti-parallel vortex lines in the initial configuration 
(without any other perturbations) the lines approach each other without deforming and finally
annihilate into soliton-like excitations which further decay into sound waves. This is the same behavior found
for a vortex--anti-vortex pair in dissipative two-dimensional superfluids, translationally extended to the 
third dimension. See \cite{Ewerz:2020wyp} for a detailed discussion of the annihilation process in the
two-dimensional holographic superfluid, and \cite{Wittmer:2020mnm} for a detailed comparison
of that two-dimensional annihilation process in holography and in dissipative Gross--Pitaevskii dynamics. 

In general, vortex reconnections occur in superfluids whenever two vortex lines come into contact 
but are not aligned anti-parallel to each other along their entire 
cores, implying that their phase structures cannot mutually annihilate.  
Instead, the interactions between the vortices cause only
segments of the vortex cores to align anti-parallel to each other, resulting in the 
annihilation of two single points, thus causing the lines to break up and form new vortex lines.
In other words, both vortex lines break up into two 
pieces which recombine with the pieces of the respective other line. 
Such a process is called reconnection.
It can generally occur for vortices of all shapes (and in principle for all winding numbers) and typically 
leads to the formation of larger vortex structures.
By definition, annihilations of vortex lines are a special type of 
reconnection where the entire cores are anti-parallel. 
Over the past decades, vortex reconnections have attracted considerable interest, mainly 
due to their huge importance in superfluid turbulence
\cite{Schwarz1985,Schwarz1988,Koplik1993,deWaele1994,Gabbay1998,Leadbeater2001,Ogawa2002,Nazarenko2003,Tebbs2010,Nemirovskii2014,Galantucci2018,Vinen2002}. 

\begin{figure}[t]
	\begin{center}
		\includegraphics[width=\linewidth]{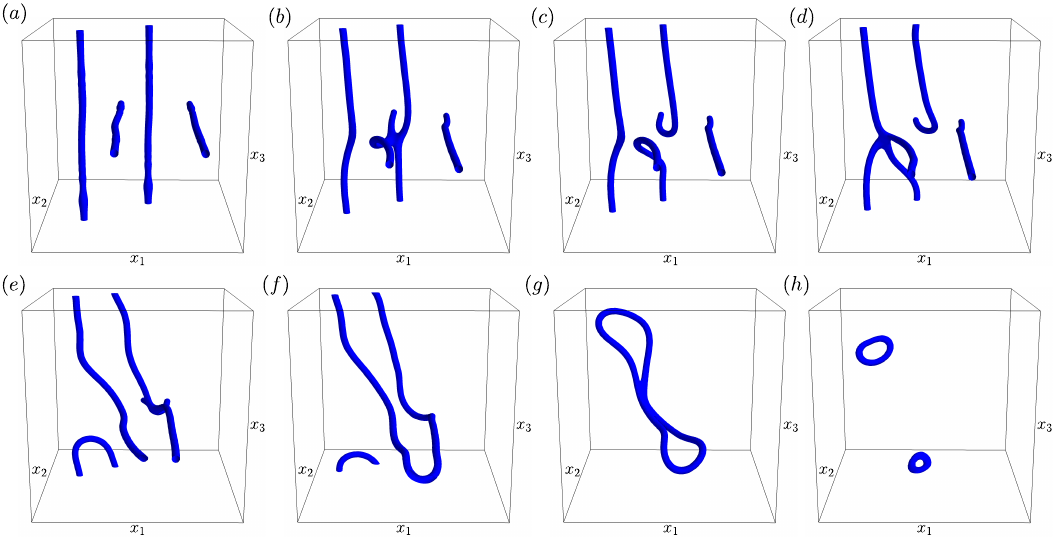}
		\caption{%
			Snapshots of isosurfaces of the superfluid density $n(t,\bm{x})$ (defining value $n/n_0=0.23$) 
			showing the time evolution of four initial vortex lines undergoing reconnections.
			The initial configuration consists of two pairs of vortex-antivortex lines aligned 
			along the $x_2$- and $x_3$-axis in such a way that the respective lines oriented along
                        the same direction are separated by a line in the other direction and therefore cannot directly annihilate. 
			Instead, the lines deform and reconnect pairwise to form new vortex lines and rings.
			Eventually, only two vortex rings remain which quickly contract and disappear due to the strong dissipation.
                        The snapshots (from (a) to (h)) are taken at $t=10, 99, 115, 135, 175, 200, 235, 275$, respectively. 
		\label{fig:Reconnections}
		}
	\end{center}
\end{figure}

To study the reconnection process in detail in the holographic superfluid 
we prepare a simple initial condition with two vortex lines of winding numbers $\pm 1$ aligned parallel to 
the $x_2$-axis and an analogous configuration aligned parallel to the $x_3$-axis on a grid of $128^3$ points. 
By placing them appropriately in the three-dimensional domain we ensure 
that the respective anti-parallel vortex lines cannot directly annihilate. 
The vortex configuration at time $t=5$, when the lines have fully developed 
their density profile, is displayed in panel $(a)$ of Fig.\ \ref{fig:Reconnections}.
%
%
As the system is propagated in time, each vortex line
moves in the flow field of the respective other vortex lines.
Since the alignment of the vortices is not translationally invariant 
the respective flow fields induce a bending of the vortex lines.
An important property of vortex defects in three-dimensional 
superfluids is that, just like for vortex rings, all segments of 
finite curvature of generic vortex lines have a self-induced velocity. 
Hence, the vortex motion induced by the flow fields is 
superimposed by their own self-induced velocity. 
In addition, the motion of the vortices is subject to 
friction with the superfluid which causes the emergence of Magnus forces.
We illustrate the time evolution in Fig.\ \ref{fig:Reconnections}. 
Note that the computational domain is periodic. 
Panel $(b)$ shows how the two `innermost' vortex 
lines effectively attract each other and come into contact. 
Evidently, in a recombination process first a short segment of each line aligns 
anti-parallel to the segment of the respective other line. 
After some time, and upon further approaching each other, the vortex lines 
break up and form new lines, \ie, they reconnect, \cf\ panels $(b)$ and $(c)$.
As time proceeds, such processes repeat, causing the vortex lines to reshuffle. 
At one point during the evolution in this example, the reconnections 
lead to the formation of one large vortex ring, see panel $(f)$.
However, since this ring is strongly elongated along one direction, it does not simply 
shrink to zero size but 
instead breaks up by another reconnection into two smaller, nearly circular rings. 
Eventually, these rings disappear by rapidly shrinking to zero size due to strong dissipation.
(For a discussion of shrinking vortex rings in dissipative GP dynamics see \cite{Berloff2007}). 
We point out that just like shrinking vortex rings or annihilating vortex lines, every 
reconnection induces the emission of rarefaction pulses and sound excitations in the superfluid. 

\begin{figure}[t]
	\begin{center}
 	\includegraphics[width=0.7\linewidth]{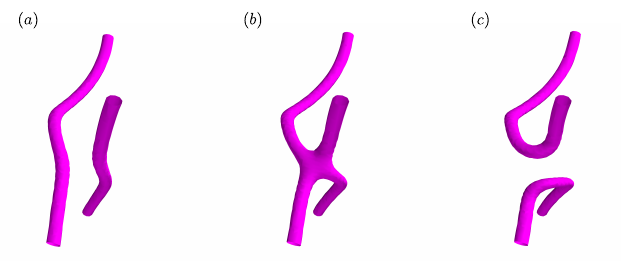}
		\caption{%
			Snapshots of isosurfaces of the superfluid density $n(t,\bm{x})$ (defining value $n/n_0=0.23$) 
                        showing a detailed view of a reconnection of two vortex lines. Only a small part of the numerical
                        domain around the reconnecting vortex line segments is shown. The snapshots (from left to right) are taken at 
                        $t=80, 102,110$, respectively. 
		\label{fig:reconnectiondetail}
		}
	\end{center}
\end{figure}

An enlarged view of the first reconnection process (\cf\ panel $(b)$ in Fig.\ \ref{fig:Reconnections}) is shown
in Fig.\ \ref{fig:reconnectiondetail}. The three panels cover a period of 30 unit timesteps around the reconnection. 

Finally, we point out that vortex reconnections are typical events in the evolution
of any sufficiently dense tangle of vortex lines, independently of
whether the underlying superfluid is dissipative or not. For the non-dissipative
case this was studied for example in \cite{Nowak:2011sk}.

\section{Bulk scalar charge density}
\label{app:bulkhcheese}
\begin{figure}[t]
	\includegraphics[width=0.3\columnwidth]{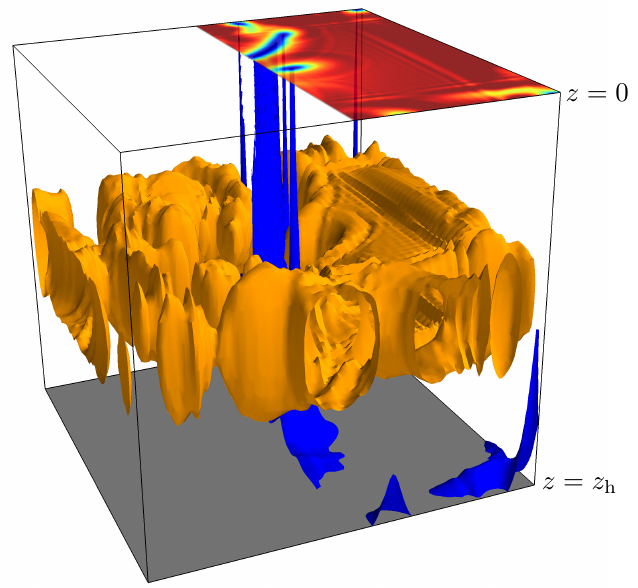}
	\includegraphics[width=0.3\columnwidth]{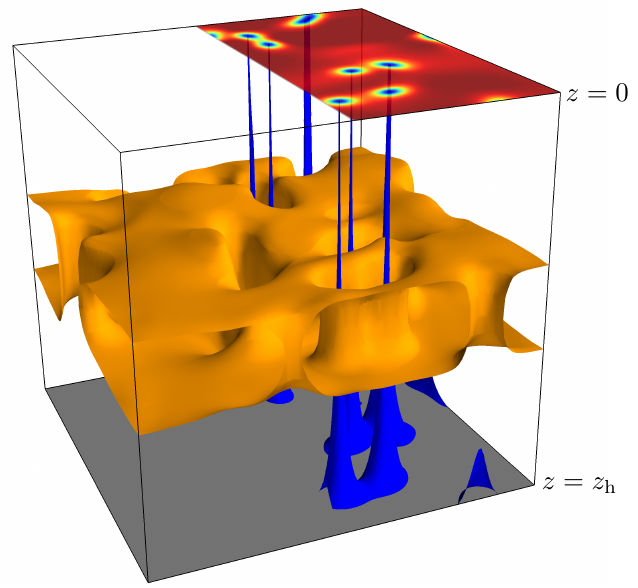}
	\includegraphics[width=0.3\columnwidth]{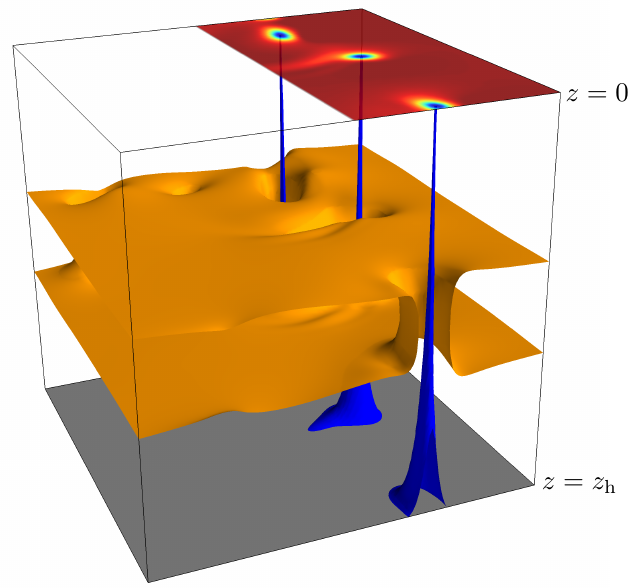}
	\caption{%
		Snapshots of isosurfaces of the bulk scalar charge density $\sqrt{-g}\,|J^0|$ (orange surfaces, defining 
		value $16$) at times $t=10$ (left panel), $t=40$ (center panel), and $t=130$ (right panel) outgoing from  
		two-dimensional slices (fixed-$x_3$ slices) of the boundary superfluid.
		For $x_1<64$, we also plot isosurfaces of the scalar field $|\Phi|^2/z^6$ (blue surfaces, defining 
		value $1.88$) and the superfluid density \mbox{$n =\lvert \psi\vert^2 $} on the boundary slice.
		Note that the domain is periodic in the spatial coordinates $\bm{x}$ of the superfluid. 
	\label{fig:CheeseEarlyAndLate}
	}
\end{figure}

%
To gain a deeper understanding of the dynamics of vortex ensembles in the 
holographic superfluid, it is interesting to study not only the 
superfluid field configuration $\psi$ itself but also its dual bulk representation. 
Here, we want to focus in particular on the bulk scalar charge density $\sqrt{-g}\,|J^0|$ which 
results from both the scalar field $\Phi$ and the gauge field $A_\mu$, \cf\ \eqref{eq:Maxwell}. 
It is sensitive not only to vortical excitations of the superfluid but also 
to rarefaction pulses and sound waves.
In Fig.\ \ref{fig:CheeseEarlyAndLate}, we 
display snapshots of the bulk-field 
configuration outgoing from fixed-$x_3$ slices of the three-dimensional 
superfluid at three times characteristic for the evolution originating from an initial ensemble of
vortex rings (initial condition of type $\mathcal{B}$). 
%
%
To illustrate the bulk-field configurations, we again plot isosurfaces of the scalar 
charge density $\sqrt{-g}\,|J^0|$ (orange surfaces, defining value $16$), on the 
entire $(x_1,x_2)$-domain corresponding to the respective choice of $x_3$-slice. 
We want to compare the characteristics of the charge density at three different
typical times of the evolution starting from a dense vortex tangle. 
We also plot isosurfaces of the scalar field $|\Phi|^2/z^6$ (blue surfaces, defining value $1.88$).
However, in order to not impede the view on the charge density, we 
restrict these isosurfaces to only half of the $(x_1,x_2)$-domain, for $x_1<64$.
We recall that the field $|\Phi|^2/z^6$ reduces to the superfluid 
density $n$ in the limit $z\to0$.
We plot $n$ on the $z=0$ slice with the same 
color map as used in Fig.\ \ref{fig:LinesCheese} in the main text.
The gray area at $z=\zh$, plotted on the entire $(x_1,x_2)$-domain,
again depicts the black-hole horizon.
(For comparison, in Fig.\ \ref{fig:LinesCheese} the defining value for the isosurfaces
of the scalar field $|\Phi|^2/z^6$ is $1.88$, while those for the isosurfaces of the scalar
charge density $\sqrt{-g}\,|J^0|$ are $12.3$ (left panel) and $16$ (right panel).) 

The left panel of Fig.\ \ref{fig:CheeseEarlyAndLate} illustrates 
that during the very early stages of the evolution of the vortex 
ensemble the scalar charge density is strongly riddled with  
holes and, in addition, strongly perturbed by many notches and smaller ripples. 
The holes are caused by vortex excitations in the dual superfluid. 
The notches and ripples, on the other hand, stem 
from dual rarefaction pulses and sound excitations, respectively. 
These are in part created by reconnection events and shrinking 
vortex rings, and in part by the early build-up process of the defects
giving rise to artifacts of the numerical initialization, see appendix \ref{app:IC}. 
Note that the blue arc in the $z=0$ plane indicates a cut through a vortex line segment 
which happens to lie in the chosen $x_3$-slice. The corresponding elongated holes
in the bulk charge density are seen here as the blue tubes emanating from the arc,
similar to the case of straight lines shown in \ Fig.\ \ref{fig:LinesCheese}. 
The center panel of Fig.\ \ref{fig:CheeseEarlyAndLate} illustrates that, at somewhat later times,
the scalar charge density is slightly less but still strongly perturbed by ripples, notches and holes. 
These perturbations are due to reconnections 
between vortex lines and vortex rings as well as due to the 
shrinking and disappearing of individual rings.  
Due to the large number of such events, many rarefaction 
pulses and sound waves are present in the system.
At this stage, the perturbations due to the initial
build-up of the defects at the start of the simulation are no longer present after having been damped out. 
At intermediate to late times, as illustrated in the right panel of 
Fig.\ \ref{fig:CheeseEarlyAndLate}, the 
ripples and notches have mostly disappeared and the charge density has smoothed out. 
Holes corresponding to a small number of vortex defects persist.
The number of rarefaction pulses and 
sound excitations has significantly decreased 
since they are produced predominantly in the now only 
rarely occurring reconnections of two vortex lines 
or rings, or the disappearing of single vortex rings. 
Nonetheless, in the aftermath of such events, they are also 
discernible in the charge density, just as at early times.   
Hence, while at early times short-wavelengths rarefaction 
pulses and sound waves are very prominent, at intermediate 
and late times their number has significantly decreased 
and they are produced only occasionally.
The general findings described here hold not only for the specific bulk views we display in 
Fig.\ \ref{fig:CheeseEarlyAndLate} but also for bulk views outgoing from all 
other two-dimensional slices of the three-dimensional superfluid. 

\end{appendix}

\end{document}